\documentclass[usegraphicx,usenatbib]{mn2e}
\usepackage{psfig}

\renewcommand\[{\begin{equation}}
\renewcommand\]{\end{equation}}

\def\Mpc{\,{\rm Mpc}}\def\kpc{\,{\rm kpc}}
\def\erg{\,{\rm erg}}\def\s{\,{\rm s}}
\def\keV{\,{\rm keV}}
\def\yr{\,{\rm yr}}\def\Myr{\,{\rm Myr}}\def\Gyr{\,{\rm Gyr}}
\def\pc{h^{-1}\,{\rm pc}}\def\K{\,{\rm K}}
\def\msun{{\rm M_{\odot}}}
\def\kms{\,{\rm km}\, {\rm s^{-1}}}
\def\pa{\partial}

\def\d{{\rm d}}\def\e{{\rm e}}

\def\i{\relax\ifmmode{\rm i}\else\char16\fi}

\def\lesssim{{_ <\atop{^\sim}}}
\def\lta{\lesssim}
\def\grtsim{{_ >\atop{^\sim}}}
\def\gta{\grtsim}

\def\fracj#1#2{{\textstyle{#1\over#2}}}

\def\lesssim{\mathrel{\hbox{\rlap{\hbox{\lower4pt\hbox{$\sim$}}}\hbox{$<$}}}}
\def\gtrsim{\mathrel{\hbox{\rlap{\hbox{\lower4pt\hbox{$\sim$}}}\hbox{$>$}}}}

\def\apj{ApJ}

\def\mnras{MNRAS}

\def\apjl{ApJL}
\def\aap{A\&A}
\def\apss{Astrophysics and Space Science}

\begin{document}

   \title[Heating cooling flows]
   {Heating cooling flows with jets}

   \author[Omma et al]
          {Henrik Omma, James Binney, Greg Bryan \& Adrianne Slyz
           \\
           Department of Physics, Keble Road, Oxford OX1 3NP\\
          }

   \date{Received July 27, 2003; accepted November 6, 2003}

   \maketitle

\begin{abstract} Active galactic nuclei are clearly heating gas in `cooling
flows'. The effectiveness and spatial distribution of the heating are
controversial. We use three-dimensional simulations on adaptive
grids to study the impact on a cooling flow of weak, subrelativistic jets.
The simulations show cavities and vortex rings as in the observations.  The
cavities are fast-expanding dynamical objects rather than buoyant bubbles as
previously modelled, but shocks still remain extremely hard to detect with
X-rays. At late times the cavities turn into overdensities that strongly
excite the cluster's g-modes. These modes damp on a long timescale. Radial
mixing is shown to be an important phenomenon, but the
jets weaken the metallicity gradient only very near the centre. The
central entropy density is modestly increased by the jets. We use a novel
algorithm to impose the jets on the simulations.
\end{abstract}


\section{Introduction}

The arrival of data from the {\it Chandra\/} and {\it XMM-Newton\/} X-ray
satellites has initiated a major change in our understanding of cooling flow
systems. In particular, the new data have convinced many astronomers that
these systems are powerfully influenced by AGN. Many aspects of this
interaction are controversial, however, because the dynamics of the
interaction between high-entropy plasma ejected by the AGN and the
surrounding thermal plasma is complex. In particular, the interaction
is inevitably time-dependent, chaotic and geometrically irregular. Numerical
simulations are capable of giving us insights into the interaction and
current thinking has been strongly influenced by the simulations of
\cite{ChurazovEtal01}, \cite{QuilisEtal01}, \cite{BruggenKaiser01},
\cite{BruggenKaiser02}, \cite{BruggenKCE02},  \cite{ReynoldsHeinzBegelman01,
ReynoldsHeinzBegelman02}, and \cite{BassonA}.  In this paper we attempt to clarify the issues
and to explore new territory with numerical simulations that are novel in
several respects. Most crucially we present the first three-dimensional
simulations that use an adaptive grid to simulate the entire cooling flow
with significant spatial resolution at the centre. We draw attention to the
possible importance of momentum input from the AGN for the dynamics of the
cluster gas, and explore a part of the parameter space that is opened up
once momentum input is considered.

In Section 2 we motivate our simulations by reviewing the history of cooling
flows and the reasons why the new data have had a profound impact. We
explain why simulations are essential for developing a sound understanding
of the connection between a cooling flow and its embedded AGN. In Section 3
we motivate our phenomenological model of how an AGN impacts on its
immediate surroundings. In Section 4 we introduce a new way to simulate jets
that are embedded in a large-scale system. Section 5 describes a pair of
simulations, and Section 6 compares these simulations with previously
published ones. In Section 7 we discuss the implications of our
work and what should be done next.

\section{Cooling-flow history}

The early X-ray satellites discovered that the deep gravitational
potential wells of clusters of galaxies and luminous elliptical
galaxies confine substantial masses of gas at the virial
temperature
\citep{Uhuru0,Uhuru2,ArielVc,Serlemitsos,Uhuru1,Einstein1,Einstein2}. It was
recognized that towards the centres of these systems the cooling
time is shorter than the Hubble time \citep{Silk76}. Motivated by
this observation, \shortcite{CowieBinney77} modelled these systems
under the assumption that they had achieved a steady state, in
which gas seeped slowly inwards as it radiated energy. The early
models did not explain what became of gas once it had cooled to
below the virial temperature and out of the X-ray band, beyond
arguing that isobaric cooling is thermally unstable, so at
sufficiently small radii clouds of cold gas might be expected to
form \citep{CowieBinney77}. \cite{FabianN} argued that these cold clouds
gave rise to the H$\alpha$ filaments that were observed
in many cooling flows \citep{Lynds70,CowieEtal83}.

As the spatial resolution of the X-ray spectroscopy increased, it
became clear that if the steady-state conjecture was valid, mass
must be dropping out of the flow into cool clouds throughout the
region of radius $r<r_{\rm cool}$, where $r_{\rm cool}$ is the
radius at which the cooling time equals the Hubble time.
\shortcite{Nulsen86} introduced a theory of a multi-component
intracluster medium (ICM) which allowed for such distributed `mass
dropout', and showed that the observed X-ray brightness profiles
of cooling flows, which were less steep than the original
steady-state models predicted, could be explained if the
distribution in temperature of the gas at a given radius were
suitably chosen.

A puzzling feature of models with distributed mass dropout was the
failure of strenuous efforts to observe objects -- ionized or
neutral clouds, or young stars -- that formed from the 10 to
$1000\msun\yr^{-1}$ of dropout that the models predicted
\citep{FabianAnRev}. Ionized gas was detected early on
\citep{Lynds70, CowieEtal83}, and molecular gas has been detected
more recently \citep{Donahue_etal00,Edge01,EdgeEtal02}, but the
mass of such matter falls short of predictions by  a factor of at
least ten, and it is more centrally concentrated than predicted.
As the quality of the available spectra improved, the models'
predictions for the X-ray spectra also came into conflict with the
data. In particular, the X-ray luminosity below $1\keV$, tended to
be significantly smaller than predicted
\citep{Stewart_etal84a,Stewart_etal84b}.

\subsection{Cracks in the edifice}

Soon after the tenth birthday of cooling-flow theory its
intellectual foundations were shot away by two papers.
\shortcite{Malagoli_etal87} showed that when gas at a range of
specific entropies is confined by a potential well with specific
entropy increasing outwards, the gas is not thermally unstable: a
region in which the temperature is lower than ambient is
under-buoyant and will sink and then oscillate around the radius
at which the ambient medium has the same specific entropy. In
principle, cooling causes these oscillations to be over-damped
\citep{BalbusSoker89, Tribble89}, but the growth rate is low and
in a real system turbulent damping is likely to suppress the
over-stability. In a second lethal attack on the steady-state
cooling-flow model, \cite{MurrayBalbus92} showed
that the differential equations that govern trapped cooling gas do
not tend to a steady state. Rather, as the system evolves from a
generic initial condition, the importance of partial derivatives
with respect to time increases secularly because the size of $\pa
x/\pa t$ is not of order $x$ over the age of the system, as
\cite{CowieBinney77} and subsequent papers had implicitly argued,
but $\sim x/(t_c-t)$, where $t_c$ is the time of a future cooling
{catastrophe}, when the local density will diverge.

In view of these theoretical developments, a handful of theorists argued
that a cooling flow is an unsteady response to heating by a galactic
nucleus, the nuclear activity being episodically stimulated by the
development of a cooling catastrophe in its environs \citep{TaborBinney93,
BinneyTabor95, Binney96, CiottiOstriker97, Binney99, CiottiOstriker01}.
Unfortunately, the steady-state cooling-flow picture had by this stage
developed such momentum that it could not be derailed by mere theoretical
considerations. A variety of mechanisms was invoked to explain the conflict
between the basic model and the new data. These included strong internal
absorption \citep{AllenFabian97}, magnetic locking of over-dense regions
\citep{Tribble89,Balbus91}, abundance anomalies \citep{FabianEtal01, MorrisFabian03},
and conduction of heat from large to small radii
\citep{BertschingerMeiksin,NarayanMedvedev01}.

\subsection{Decisive new data}

The arrival of data from Chandra and XMM-Newton has finally turned the tide
of opinion away from the distributed mass-dropout hypothesis. In particular, the new
data show that internal absorption is not viable as an explanation of the
paucity of flux below $1\,$keV because it suppresses very soft photons that
are observed in abundance
\citep{BohringerEtal01}.  Moreover, spatially resolved spectra are
incompatible with the predicted multiphase ICM: there is no evidence for
plasma at more than one temperature at a given location
\citep{TamuraEtal01,PetersonEtal02}. Within the cooling
radius the temperature of the ICM is found to decline with decreasing
radius, but has a floor value that varies from system to system. The floor
almost always lies above $1\,$keV, so the paucity of flux below $1\,$keV
simply reflects an absence of very cool gas, in clear conflict with the
steady-state model. \shortcite{KaiserBinney03} have shown, by contrast, that
if cooling flows are episodically heated, very cool gas will rarely be seen,
because in any given system it appears only fleetingly and in small quantity.

In addition to completing the falsification of the mass-dropout model, the
new data have provided morphological evidence for heating by AGN. It has
long been recognized that cooling flows generally have embedded non-thermal
radio sources \citep{FabbianoGT}, but it used to be argued that there was no
convincing evidence that these sources had a significant impact on their
cooling flows \citep{AllenEtal01,FabianEtal01}, even though the beautiful
radio map of Virgo A by \shortcite{OwenEilekKassim00} showed that in Virgo
AGN-powered ultrarelativistic electrons densely permeate the region
$r<r_{\rm cool}/2$.  Imaging data for several systems now show `cavities' in
the X-ray emitting plasma that coincide to some degree with peaks in the
non-thermal radio emission
\citep{BohringerEtal93,ChurazovEtal00,McNamaraEtal00,McNamaraEtal01,BlantonEtal01}.
The interpretation that the cavities are `bubbles' inflated by radio sources
in the thermal plasma, has gained currency
\citep{ChurazovEtal01,QuilisEtal01,BruggenKaiser01,BruggenKaiser02,BruggenKCE02}.
Consequently, it is now widely believed that cooling-flow gas is
significantly affected by AGN.

\section{Key Questions}

Many fundamental questions regarding the nature of AGN/cooling-flow
interaction remain to be answered. For example:

\begin{enumerate}

\item Are the observed cavities inflated by ultrarelativistic jets
or by sub-relativistic bipolar flows, or by a combination of the
two? We shall see that the answer to this question has a strong
bearing on the dynamics of the cavities: an ultrarelativistic jet
carries very little momentum, so the cavities it inflates simply
rise buoyantly, as in the simulations of \cite{ChurazovEtal01},
\cite{QuilisEtal01}, \cite{BruggenKaiser02} and others; a
subrelativistic wind imparts momentum as well as energy to the
cavity, which is thus driven up through the ambient gas, with
implications for the stability and longevity of the cavity.

\item How is the energy that we know lies  within the observed cavities,
transferred to the thermal X-ray emitting gas?

\item Do the observed cavities in cooling flows such as those in Hydra and
Perseus heat the systems fast enough to offset radiative cooling?
While it is now clear that cooling flows are heated by AGN and that
locally partial derivatives with respect to time are significant terms in
the governing equations, it is not clear whether the azimuthally averaged
profiles of flows are strongly time dependent. At one extreme
\cite{KaiserBinney03} present a picture in which heating is unimportant for
several hundred megayears between catastrophic AGN outbursts, and the radial
density profile evolves significantly between outbursts. At the other
extreme there is the possibility that the cores are in a turbulent
steady-state in which azimuthally-averaged heating and cooling are in
balance \citep{TaborBinney93}.

\item On what timescale does a cooling-flow settle
to approximate hydrostatic equilibrium after a nuclear outburst?

\item
How visible in X-rays will be non-equilibrium structures, such as cavities
and shocks, that the AGN generates?

\item
What is the radial distribution of injected energy after approximate
hydrostatic equilibrium has been restored? This distribution effectively
determines the predicted X-ray brightness profile and the time elapse between
outbursts.

\item
What impact does AGN-induced radial mixing have on metallicity gradients? Is
this impact compatible with models of metal enrichment and observed
metallicity distributions?

\item
We must expect the outflow from the AGN to fluctuate on very short
timescales.  What observable phenomena will be generated by these
fluctuations? A priori we expect the shortest timescale fluctuations to be
evident at the smallest radii. What phenomena  are expected at different
radii, and what are the associated timescales?

\item Jets from accreting objects are known to precess --  most famously in
the case of SS 433 \citep{Milgrom79}. How would the inflation of cavities be
affected by jet precession or by a wide opening angle of the jet \citep{Soker03}?

\item Gas trapped in a potential well is likely to have non-zero
angular momentum. As the gas cools, the dynamical importance of
angular momentum is liable to increase -- in proto-spiral galaxies
it evidently becomes dominant. Radial stirring of gas by an AGN is
liable to move angular momentum outwards and make it dynamically
less important. How rapid is such angular-momentum transport? What
impact does it have on the evident failure of giant elliptical
galaxies to form disks recently \citep{EmsellemSAURON02}? and on
the location of the molecular gas that has been found in cooling
flows \citep{Donahue_etal00,Edge01,EdgeEtal02}?

\end{enumerate}

Since the interaction between an AGN and a cooling flow is inherently
non-spherical and unsteady, simulations have a vital role to play in
answering these questions. Several sets of simulations of AGN / cooling-flow
interaction have appeared in the literature \citep{ChurazovEtal01,
QuilisEtal01, BruggenKaiser01, BruggenKaiser02, ReynoldsHeinzBegelman01,
ReynoldsHeinzBegelman02, BruggenKCE02, BassonA}.  For reasons of
computational cost, most published simulations assume spherical or
axisymmetric symmetry or use a rather coarse spatial resolution. We present
simulations that are fully three-dimensional and achieve competitive spatial
resolution in a large box by exploiting adaptive grids.

\subsection{Heating mechanism}

Published simulations adopt a variety of approaches to
the way in which the AGN transfers energy to its environs.

\shortcite{CiottiOstriker97,CiottiOstriker01} consider
inverse-Compton heating of the gas. This is an inefficient process
since a photon of energy $E$ transfers to the gas a fraction $\sim
E/m_ec^2$ of its energy, so even if the Thompson-Compton optical
depth to the AGN is significant, radiation carries away most of
the AGN's output unless the spectrum of the radiation is hard.
Moreover, for observed cooling flows the Thompson-Compton optical depth from
infinity to radius $r$ is small down to the smallest radii ($r\gta100\,$pc)
that can be resolved in state-of-the-art simulations. Consequently, any
inverse-Compton heating of the gas is likely to be concentrated at
unresolved radii, and to impact on the simulated region by forcing an
inflow across the simulation's inner boundary. In this respect the
inverse-Compton model is similar to the model on which we shall concentrate.
The essential difference is that we shall assume a strongly bipolar outflow,
whereas inverse-Compton heating might produce a nearly spherical flow at the
inner boundary.

Our simulations assume that a cooling flow is primarily heated by a bipolar
flow. This assumption is based on several considerations. First it is now
generally accepted that AGN are powered by accretion onto a massive black
hole. Bipolar outflows are observed around most accreting objects, be they
forming stars \citep{ArceG}, accreting black holes that power AGN \citep{PoundsEtal03}
and micro-quasars \citep{MirabelR99}, or forming galaxies \citep{PettiniEtal02}. Hence, it is
natural to expect a bipolar outflow to emerge from an AGN that is heating a
cooling flow.

A second reason to believe that cooling flows are heated by bipolar flows
follows from the observation \citep{FabianCanizares88} that the apparent
luminosities of many galactic nuclei are several orders of magnitude smaller
than is predicted by Bondi-Hoyle accretion onto the black holes that are
known to reside there, given plausible lower limits on the central gas
density in the cooling flow. The Advection Dominated Accretion Flow (ADAF)
model of accretion onto black holes has been extensively explored as a way
of explaining this surprising result \citep{Ichimaru,NarayanYi}. However,
the basic physical premise of the ADAF model, that coulomb scattering is
the dominant process for establishing equipartition between electrons and
ions, is probably false \citep{binney03}.  \shortcite{BlandfordBegelman99}
propose an alternative to the ADAF model as an explanation of the low
luminosities of galactic nuclei. In their ADIOS model most of the accretion
energy that is released as plasma falls into the black hole is used to drive
a wind from the surface of the accretion disk. Most of the gas that falls
onto the outer edge of the accretion disk is carried by this wind away from
the black hole, with the result that the hole's accretion rate is much smaller
than the disk's accretion rate.

Our simulations assume that the ADIOS model correctly predicts
that much of the energy released by accretion onto the black hole
is channelled into a sub-relativistic bipolar flow. We emphasize
that this flow is probably distinct from the highly relativistic
flow that generates jets of synchrotron radiation in many well
observed systems, such as M87. Key differences between the bipolar
flow on which we focus and the underlying synchrotron jets include
(i) the former is subrelativistic since material is blown off the
accretion disk at the local Kepler speed, while superluminal
motion and one-sidedness show the latter to have a Lorentz factor
of several; (ii) the former  will comprise an ordinary H/He gas,
while the latter is quite possibly an e$_\pm$ plasma; (iii) the
former is a direct and inevitable corollary of accretion while the
latter probably draws its energy from the black hole's spin via
vacuum breakdown \citep{BlandfordZ}, and it may switch on and off
in an erratic way.

The distinction between the bipolar flow from the disk and the one that
generates synchrotron jets is important because the latter flow probably
dominates radio maps at all radii through its superior ability to generate
ultrarelativistic electrons. However, in many observed systems both flows
are likely to be present simultaneously, so we should imagine the outflow to consist of a
series of concentric cylinders (or cones in the case of non-negligible jet
opening angle), each cylinder moving parallel to the axis at
a speed that increases from $\sim100\kms$ on the outside to $\sim c$ at the
centre. Instabilities powered by the shear within this system will steadily
transfer momentum and energy outwards, and cause ambient plasma to be
entrained at the edge. A very basic point, but one that is often overlooked,
is that the mechanical luminosity of the entire jet is orders of magnitude
greater than the synchrotron luminosity. The ratio of these luminosities
will be especially large if there is no ultrarelativistic jet at the core
of the subrelativistic outflow.

Outflows with speeds $\sim0.1c$ and mechanical luminosities of order the
Eddington luminosity are directly observed in spectra of accreting
relativistic objects \citep{PoundsEtal03}. While these outflows are suggestive,
they are not directly relevant to the systems of interest here because they
are observed in objects with comparable photon luminosities and they may
be radiatively driven \citep{KingPounds}. It is likely that in very
dense environments much of the accretion energy is degraded into photons,
while in systems with very hot and therefore rarefied central gas, the
accretion energy emerges in largely mechanical form.

\section{Simulation technique}

Even with the best current soft- and hard-ware it is impracticable
to simulate flows from the sub-parsec scales on which jets form,
out to beyond $r_{\rm cool}\sim100\kpc$. In practice one must
start at some smallest resolved scale $r_{\rm min}\simeq1\kpc$,
using a model of the jets on that scale that derives from a
mixture of physical intuition and a critical examination of
observed jets. From observations and modelling of jet evolution at
$r\simeq r_{\rm min}$, one hopes to infer the structure that the
jets must have on much smaller scales if they are to have the
inferred structure of scales $r_{\rm min}$ and above.

\subsection{Generating jets}

On a given scale, a jet is characterized by the rates $\dot m$,
$\dot P$ and $\dot E$ at which it injects mass, momentum and
energy into the larger-scale plasma. If  the
characteristic speed of flow within the jet, $v_{\rm jet}$, is
subrelativistic,  these three numbers
are clearly related by $\dot P=\dot mv_{\rm jet}$ and $\dot E=\dot
m(u+\fracj12v_{\rm jet}^2)$, where $u$ is the specific internal
energy of the jet material. So long as the jet is highly
supersonic, $u\ll\fracj12v_{\rm jet}^2$. Entrainment causes $\dot
m$ to rise and $v_{\rm jet}$ to decline as one proceeds down the
jet, such that $\dot P$ remains approximately constant.
Entrainment leads to an increase in $u$ relative to the kinetic
contribution to $\dot E$, while radiative cooling causes $u$ to
decrease.

Our simulations employ the three-dimensional adaptive-mesh
hydrocode ENZO \citep{BryanNorman97} with a Piecewise Parabolic
Method (PPM) Riemann solver \citep{ColellaWoodwardPPM84}.  The
refinements are dynamically generated based on the density
gradient down to 6 levels of refinement, with an effective
resolution of $1024^3$ cells. The computational box is $635\kpc$
on a side, and as a result the central $\sim35\kpc$ is covered by
cells measuring $620\pc$.  Periodic boundary conditions are
enforced.

We simulate the action of a pair of sub-grid jets by adding mass,
$x$-momentum and energy to cells that lie within the $yz$ plane at
$x=\pm\epsilon$, where $\epsilon\le1\kpc$. During a timestep of
length $\delta t$, the injected mass $\dot m\delta t$ is
distributed over cells that lie in a thin disk according to the
window function
 \[
w(r)\propto\cases{\e^{-r^2/2r_{\rm jet}^2}&for $r<r_{\rm jet}$,\cr0&otherwise.}
\]
 Here $r=\sqrt{y^2+z^2}$ is the distance of the cell from the $x$-axis and
$r_{\rm jet}=\epsilon/0.3$. The injected momentum and energy are
distributed in the same manner. We present results for two values
of $r_{\rm jet}$, namely 2 and $3\kpc$ to illustrate the effect of
changing this parameter.  The cross-sectional area of the jet at
the highest resolution is resolved into $\sim35$ ($78$) cells for
the $r_{\rm jet}=2\kpc$ ($3\kpc$) case.

The first increments in the dynamical variables after a jet is switched on
increase the mean velocity of material in a disk cell only very slightly
because the injected momentum has to be shared with a substantial quantity
of stationary gas.  Consequently, little of the injected energy is tied up
in the kinetic energy of the cell. Hence, nearly all the injected energy is
used to heat the gas that was originally in the cell, and this gas expands
in response. The resulting decline in the density of gas in disk cells
causes subsequent momentum increments to yield ever larger bulk velocities
for disk cells, and gradually a significant fraction of the injected energy
goes into bulk kinetic energy.

From this discussion it will be seen that the speed $v_{\rm jet}$
that characterizes the ratio of the increments in momentum and
energy, controls the relative importance of heating, which tends
to expand the nuclear gas isotropically, and bulk motion, which
generates bipolarity within the cooling flow. In this paper we
restrict ourselves to simulations in which $v_{\rm
jet}=10\,000\kms$ and $\dot m=2\msun\yr^{-1}$, which corresponds
to a total power $P=6\times10^{43}\erg\s^{-1}$ for two jets. The
jets remain powered for $100\Myr$ so that they inject
$2\times10^{59}\erg$ in total. We shall explore the effects of
changing the jet power in a later paper. The low value of $v_{\rm
jet}$ adopted here contrasts with the much higher jet velocities ($\sim100\,000\kms$)
studied in
\cite{ReynoldsHeinzBegelman01,ReynoldsHeinzBegelman02} and
\cite{BassonA}.

Our approach to jet simulation avoids the imposition of boundary conditions
at points internal to the simulation. It involves adding source terms for
each cell to generate the appropriate $\dot m$, $\dot P$, and $\dot E$.
Given these, the code computes increments in the internal energy $u$.
Neither the temperature nor the density at the base of the jet is
hard-wired; they evolve as natural consequences of $\dot m$, $\dot P$, and
$\dot E$.  Our approach guarantees a gradual dynamical response to the jets
being either turned on or off.

The power of our jets is smaller than the X-ray
luminosities of the cluster that we model
($L_X=2.5\times10^{44}\erg\s^{-1}$ \citep{DavidEtal00}), so jets of this power could
significantly modify the cooling flow only if they were active for
much of the time. It seems likely that jet activity is
intermittent, and our goal in this paper is to understand the
impact of a single outburst. The relevant dynamical timescale of
the cluster gas turns out to be quite long (see below), so the
dynamical impact of an outburst is evident more than a gigayear after
the jets have switched off. By contrast, in the absence of a
further outburst, the core will proceed to a cooling catastrophe
within $\sim300\Myr$ (e.g.\ \cite{KaiserBinney03}). Hence if we are
to understand the impact of a single outburst, we must suppress
radiative cooling. We reserve to a later paper simulations that
include radiative cooling and repeated outbursts.

\subsection{The cluster model}

\begin{figure}
\psfig{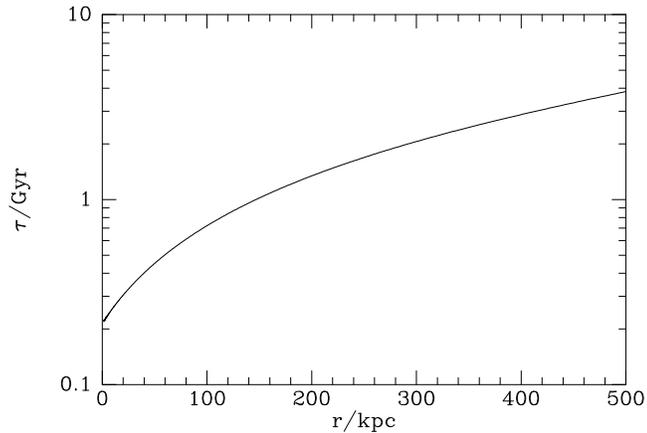} \caption{The period of small
radial oscillations as a function of radius in the initial
model.\label{BVfig}}
\end{figure}

The background that is disturbed by the jets is based on {\it
Chandra\/} observations of the Hydra cluster by
\shortcite{DavidEtal00}. The gravitational potential is that of
the NFW model that David et al.\ fitted to their data. From this
potential we calculated electron densities for an isothermal
distribution of plasma, and chose the temperature
$T=3.7\times10^7\,$K that provides the best fit to the electron
density that David et al.\ inferred. The equations of motion were
integrated for $100\Myr$ before the jets were switched on to
ensure that the system was not far from numerical equilibrium when
the jets were ignited. The jets fired for $100\Myr$ before turning
off, and the subsequent evolution was followed for $2.6\Gyr$.

\begin{figure}
\centerline{\psfig{file=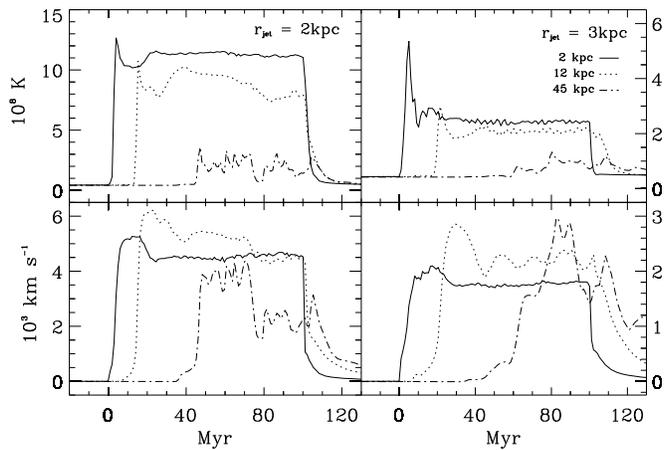,width=1.07\hsize}}
\caption{Temperature and $x$-velocity as a function of time for
points lying on the jet axis at $x=2$, $12$ and
$45\kpc$.\label{axisTVfig}}
\end{figure}

\begin{figure}
\centerline{\psfig{file=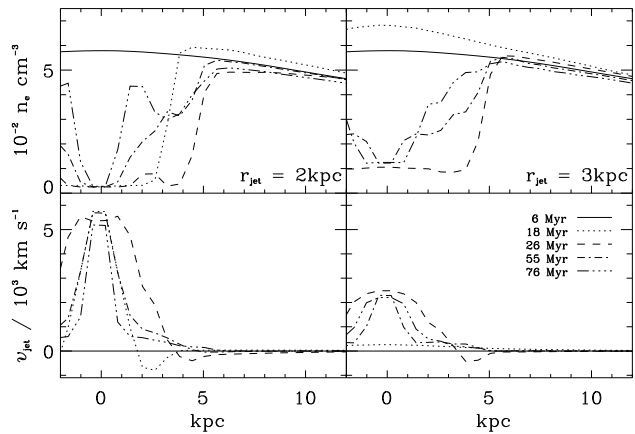,width=\hsize}}
\caption{Electron density and the $x$ component of velocity as
functions of perpendicular distance from the jet axis sampled at
$x=12\kpc$. Figs.~\ref{densplots2kpc} and \ref{densplots3kpc} show
two-dimensional slices through the density field at each of the
times plotted here.\label{crossfig}}
\end{figure}

The jets deform the initially spherical surfaces of constant
specific entropy.  After the jets have turned off, the surfaces of
constant specific entropy are expected to wiggle in and out with a
characteristic period that increases from small radii to large.
The  period of these oscillations are of order
$\tau=2\pi/\omega$, where the Brunt-Vaisala frequency
$\omega$ is given by
 \[
\omega^2={\d\Phi\over\d r}{\d\ln\sigma\over\d r},
\]
 where $\Phi$ is the gravitational potential and $\sigma=P\rho^{-\gamma}$ is
the entropy index. In Fig.~\ref{BVfig} we plot $\tau$ as a function of
radius in our initial model of the Hydra cluster. The period rises from
$\sim0.2\Gyr$ at the origin to $1\Gyr$ at $150\kpc$. Hence the period is
everywhere longer than the likely lifetime of a jet.

\section{Results}

\begin{figure*}
\centerline{\psfig{file=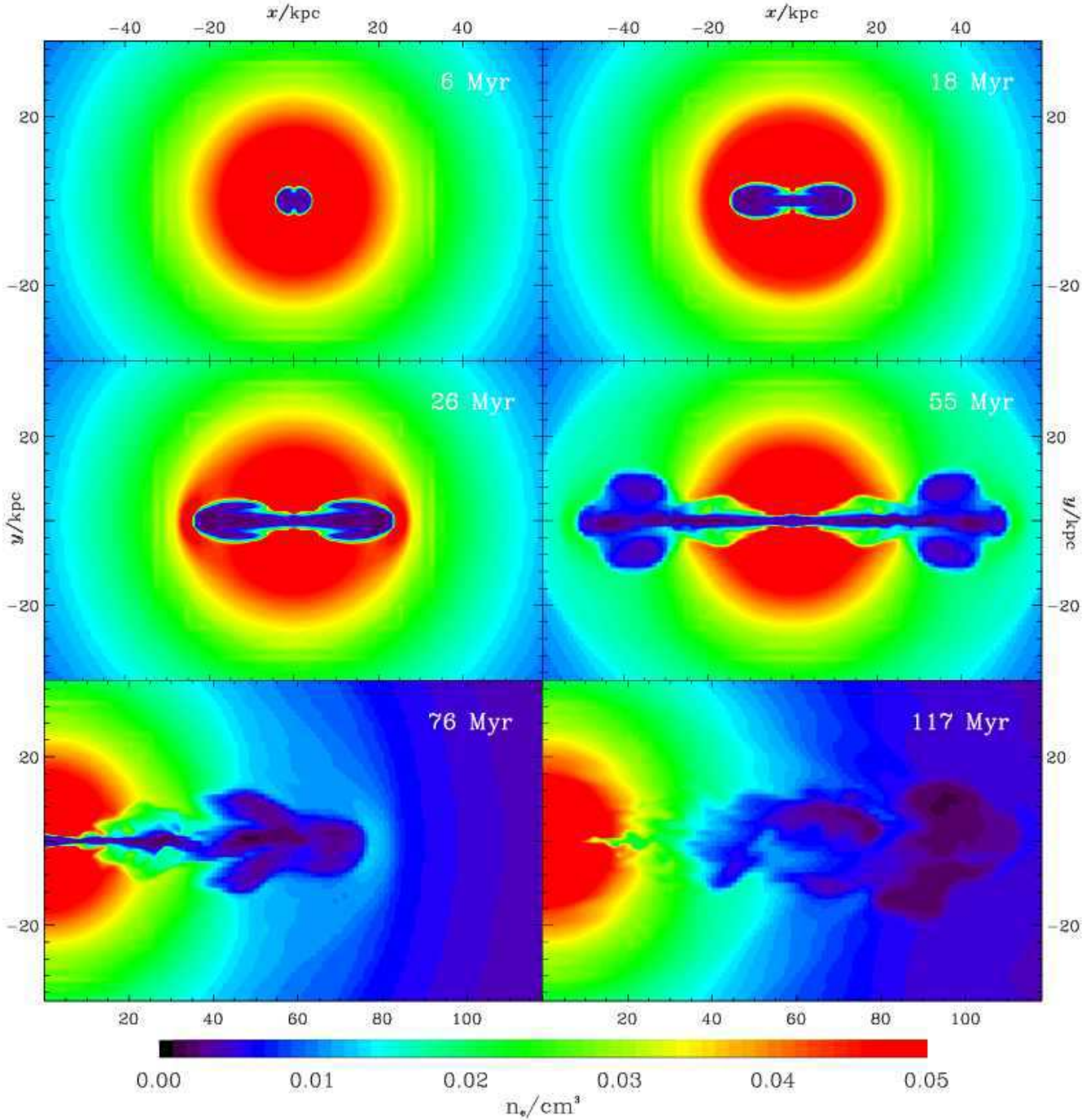,width=\hsize}} \caption{Density
in the plane $z=0$ at $6$ to $117\Myr$ after the jets turn on, in
the case of the $2\kpc$ jets.  The intensity scaling and length
scale is the same for each image, but the last two panes show only
the positive $x$-hemisphere. \label{densplots2kpc}}
\end{figure*}

\begin{figure*}
\centerline{\psfig{file=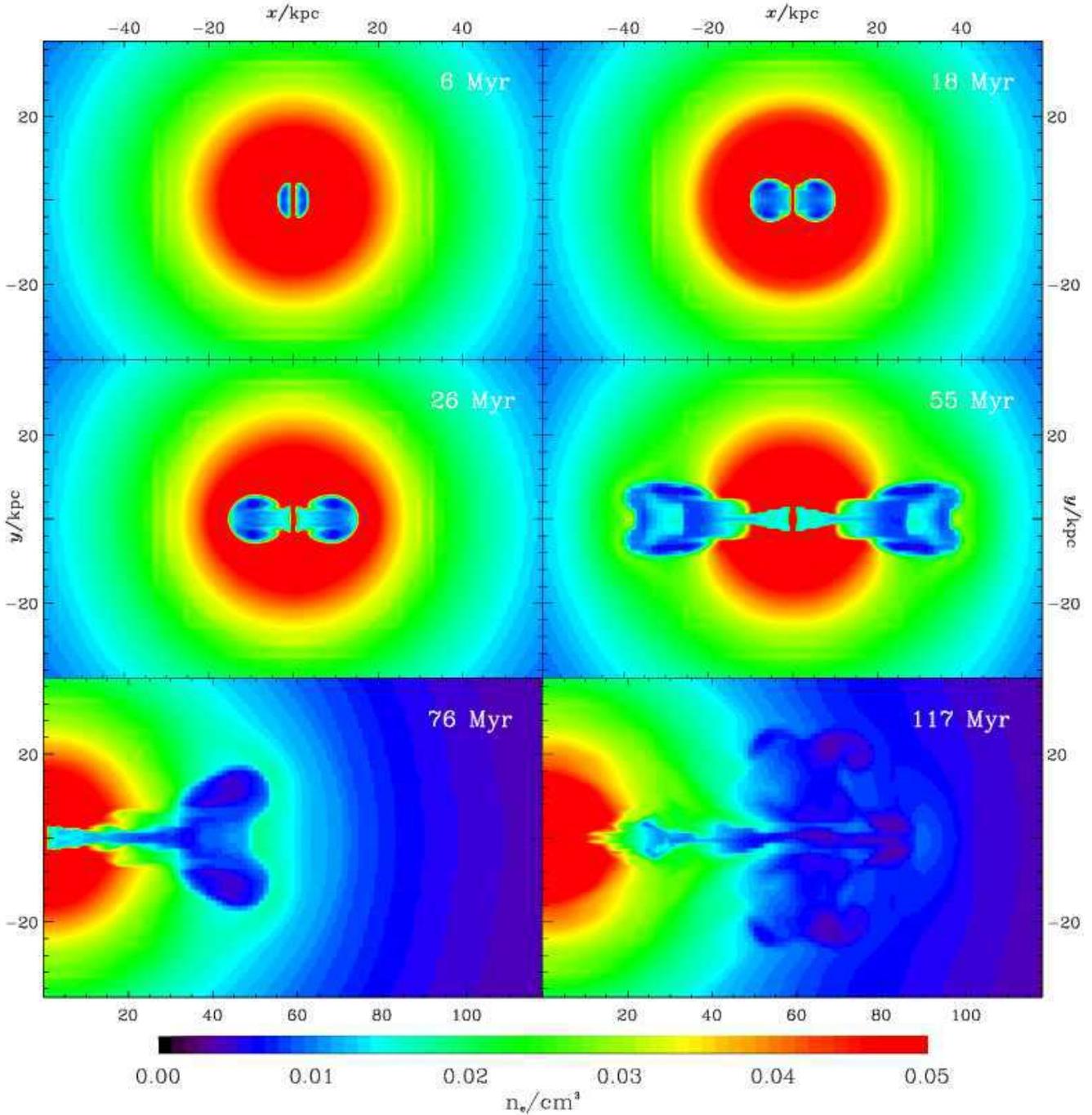,width=\hsize}} \caption{Same
images as in Fig.~\ref{densplots2kpc}, but for the case of the
3$\kpc$ jets. \label{densplots3kpc}}
\end{figure*}

Fig.~\ref{axisTVfig} describes the process of jet formation once
mass, momentum and energy start being injected into the disk by
plotting the temperature and $x$-velocity on the jet axis at $2$,
$12$ and $45\kpc$ from the cluster centre. The upper full curves
show that the temperature at $x=2\kpc$ rises steeply to a sharp
peak as injection starts. The temperature then fluctuates before
settling to a plateau that lies above $10^9\K$ in the case of the
narrower jet, and slightly below $3\times10^8\K$ in the wider
case.  The dotted curves in the temperature panels show that after
a delay this process is repeated at slightly lower temperatures at
$x=12\kpc$. The full curves for the $x$-velocity in
Fig.~\ref{axisTVfig} show that at $x=2\kpc$ the velocity rises
slightly less rapidly than the temperature before falling and
levelling off at a $4500\kms$ plateau for the narrow jet, and
$1800\kms$ for the wider one. Since the speed of sound in the ambient medium
is $920\kms$, the  corresponding Mach numbers with respect to the ambient medium are
$M=5$ and $M=2$. The jets' internal Mach numbers are much less than unity.
For both jets the peak velocities
are larger at $x=12$ than at $2\kpc$. In light of the lower
temperatures at $12\kpc$, this suggests that between these two
points the jet has narrowed and accelerated as in a nozzle. Thus these are
pressure-confined jets such as are thought to exist in low-power (FR I)
radio galaxies, rather than the ballistic jets of FR II sources.

The dot-dashed curves in Fig.~\ref{axisTVfig} show similar trends at
$x=45\kpc$ except that (i) the impact of the jet is delayed by $\sim45\Myr$,
and (ii) the temperature attains much smaller peak values. The lower
temperatures again suggests adiabatic expansion.  The pressure at a given
radius should be approximately time-independent since it is determined by
the ambient cooling flow. Hence, in the absence of entrainment, the
Bernoulli function of each fluid element would be conserved. Actually the
decrease in $T$ is not associated with a corresponding increase in velocity,
so entrainment probably is important. Direct measurements of mass
flux at various radii confirm this inference. At all three locations both
temperature and velocity drop precipitately soon after injection ceases.
Interestingly, the delay between the velocity dropping from location to
location is smaller than the corresponding delay that accompanied jet
ignition. Thus it takes only $15-20\Myr$ for the jet channel to fill up out
to $45\kpc$, compared to the $\sim45\Myr$ required to cut the channel that
far.

The fact that the plateau velocities at $x=2\kpc$ in
Fig.~\ref{axisTVfig} are substantially smaller than $v_{\rm jet}$
reflects the importance of entrainment at the base of the jet.
Near the plane $x=0$, gas is sucked in towards the $x$ axis, flows
through the injection disks, and is then blasted out along the
axis. This flow diminishes the final jet velocity in two ways: (i)
it diminishes the pressure just behind the injection disk, which
tends to throttle the flow; (ii) it causes the momentum injected
on the disk to be shared by a larger quantity of gas. The flow
induced through the back of the injection disk is larger in the
case of the wider jet, because the area of its disk is $2.25$
times that of the narrower disk. In fact, the plateau velocity of
the wider jet is smaller than that of the narrow jet by a very
similar factor, $2.5$.  The detail of this entrainment process is
highly artificial, but the general physical principle is sound:
gas will be entrained by a jet that has propagated out from the
AGN to the radius at which we have placed our injection disks.  By
measuring the flux of gas at $2\kpc$ we find that the $2\kpc$ jet
entrains $\sim 3\msun\yr^{-1}$ at its base, while the $3\kpc$ jet
entrains $\sim 10\msun\yr^{-1}$.

Fig.~\ref{crossfig} shows for five times electron density and $x$-velocity
along a cross-sections through the jets at $x=12\kpc$. (The density fields
in a two dimensional plane at these times are depicted in
Figs.~\ref{densplots2kpc} and ~\ref{densplots3kpc}.) The upper panels in
Fig.~\ref{crossfig} show that by $26\Myr$, injection has drastically lowered
the density on the axis, by a factor $\sim25$ in the case $r_{\rm
jet}=2\kpc$ and by a factor $\sim6$ in the case of the wider jet.  The
situation at $18\Myr$ is more complex: in the narrower jet the density is
already low out to $\sim4\kpc$ of the jet axis, but it is slightly raised in
the region beyond 4 kpc by the jet-driven flow of material away from the
centre; at this early time the wider jet has yet to reach $x=12\kpc$ so the
density lies above its original value throughout the cross section and, at
$300\kms$, the velocity is still subsonic.  At $18\Myr$ the narrower jet is
highly supersonic (with respect to the ambient medium; $v\sim6000\kms$) and
shows a significant backflow in the region $y=2-5\kpc$, with speeds up to
$970\kms$. The wider jet develops a backflow later at $x=12\kpc$, and in
both cases the backflow region moves away from $x=12\kpc$ at later times.
It is interesting that both the density and velocity plots show that the
jets become narrower with time. This fact reflects the movement downstream
of the backflow region and suggests that the artificial shear viscosity in
ENZO is rather small. The velocity profiles for $t=55$ and $76\Myr$ show
quite wide wings either side of the narrowing core. Although the velocities
associated with these wings are modest, the corresponding net momentum is a
significant fraction of the jet's momentum because the areas and densities
associated with the wings are large. Indeed, the directly measured mass flux
in the region $2<x/\kpc<12$ of either jet reveals a further mass loading of
$8-10\msun\yr^{-1}$.

Figs.~\ref{densplots2kpc} and \ref{densplots3kpc} show the density
field in a two dimensional plane ($z=0$) at the same six times for
the $2$ and $3\kpc$ jets, respectively. At the first time shown,
i.e. $6\Myr$ after the jets were turned on, two lunes of low
density can be seen, one on each side of the injection disk. These
lunes rapidly swell parallel to the $x$ axis and develop internal
structure that makes them resemble mushrooms, those for the wider
jet being shorter and plumper. By $55\Myr$ more complex structures
have developed that differ significantly between the jets: the
narrower jet has a butterfly-shaped cavity about $15\kpc$ long
that has its head about $50\kpc$ from the origin and a wide tail
around a long straight section of jet that reaches $30\kpc$ back
to the origin. By contrast, the cavity generated by the wider jet
extends only to $x\sim40\kpc$ along the axis and is shaped rather
like a red pepper. A shorter, wider jet leads back from this
structure to the origin. The structures seen at $t=76\Myr$ show
similar differences between the two jets. At $117\Myr$, $17\Myr$
after the jets turned off, the channels along which the jets ran
are rapidly filling near the origin and the outer structures are
becoming more turbulent. The heads of the cavities continue to
plough outwards. About $100\Myr$ after the jets have turned off
(not shown) the outward crashing structures undergo a fundamental
change: a stream of colder than ambient gas surges up from behind
the cavity, largely filling it and bursting out of its leading
edge. In this way the cavities turn into overdensities.

In the $76\Myr$ snapshot of the $2\kpc$ jet we see that the flow
is disturbed in the region $25<x/\kpc<35$, where the jet no longer
follows a linear path. This is a highly turbulent region with
strong internal shocks that disrupts the jet stem and almost
separates it from the established cavity. There is no analogous
feature in the $3\kpc$ jet, but do we find that such features play
an important role in simulations with faster jets.

\begin{figure}
\centerline{\psfig{file=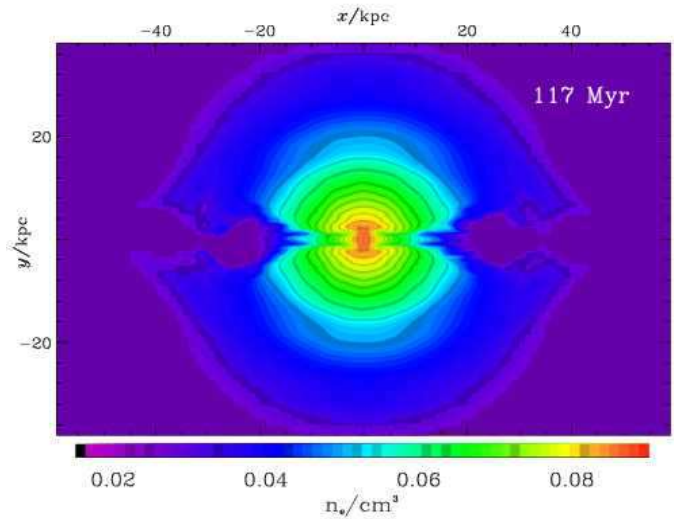,width=1.07\hsize}}
\caption{Density as shown in the $t=117\Myr$ panel of
Fig.~\ref{densplots3kpc}, but with an altered intensity scaling
which enhances the equidensity surfaces.\label{blowup}}
\end{figure}

\begin{figure}
\psfig{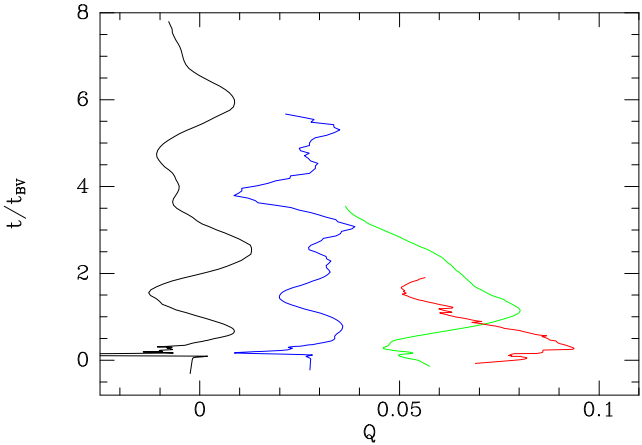} \caption{Quadrupole moment
of the density distribution as a function of $t/t_{\rm BV}$ with
$t<2.6\Gyr$ at $r=25,$ $50$, $110$ and $200\kpc$. Curves for
successive radii have been shifted horizontally by
$0.03$\label{equidens}}
\end{figure}

Fig.~\ref{blowup}, which is a centered view of the cluster core at
$t=117\Myr$, shows that the equidensity surfaces remain markedly
aspherical some time after the stem has retreated from a given
radius.  They soon after become roughly elliptical and distort
under radial oscillations that are analogous to the g-modes of a
star. Fig.~\ref{equidens} quantifies the relaxation of the
equal-entropy surfaces by plotting as a function of scaled time
$t/\tau$, where $\tau$ is the local Brunt Vaisala time from
Fig.~\ref{BVfig}, the quadrupole moment of the entropy
distribution
 \[
Q(r,t)=\int\d\theta\,\sin\theta\,{3\cos^2\theta-1\over2}\,
{\sigma(r,\theta,t)\over\overline{\sigma}(r,t)},
\]
 where $\theta$ is colatitude with respect to the jet axis and
$\overline{\sigma}(r,t)$ is the average value of the entropy index $\sigma$
on a sphere of radius $r$. At $r=25$ and $50\kpc$ a large excursion in $Q$
is evident as the jet first impacts, followed by an oscillation pattern that
looks like an interference pattern between the first few harmonics with
fundamental period comparable to $\tau(r)$. At $r=110$ and $200\kpc$ very
few cycles are seen even though the baseline in time extends to $2.6\Gyr$.

\begin{figure}
\centerline{\psfig{file=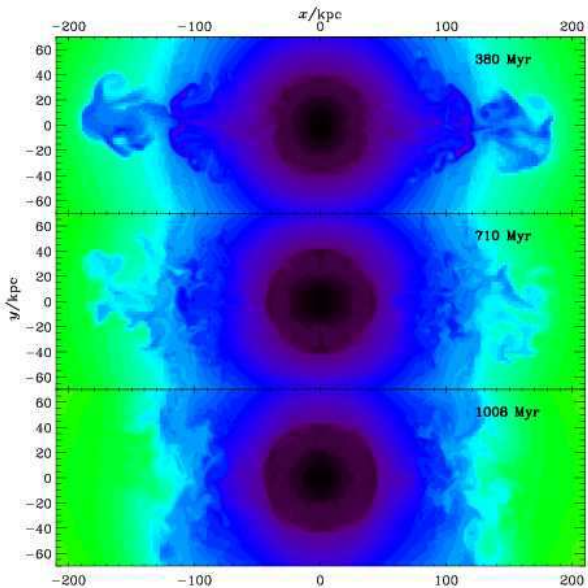,width=.96\hsize}}
\caption{Maps of specific entropy in the  plane $z=0$ at late times.\label{LateTimes3kpc}}
\end{figure}

Fig.~\ref{LateTimes3kpc} shows the entropy field for the $3\kpc$
jet at three late times. The panel for $380\Myr$ shows the flow
after the cavities have transformed into over-densities. In the
panel for $710\Myr$ we see that around $x=70\kpc$ overdense
material has fallen back and indented the otherwise fairly smooth
isoentropy surfaces at that radius. In the bottom panel we see the
same process happening further out at $1\Gyr$. This outward-moving
wave of material falling back strongly excites the system's
g-modes.

\begin{figure*}
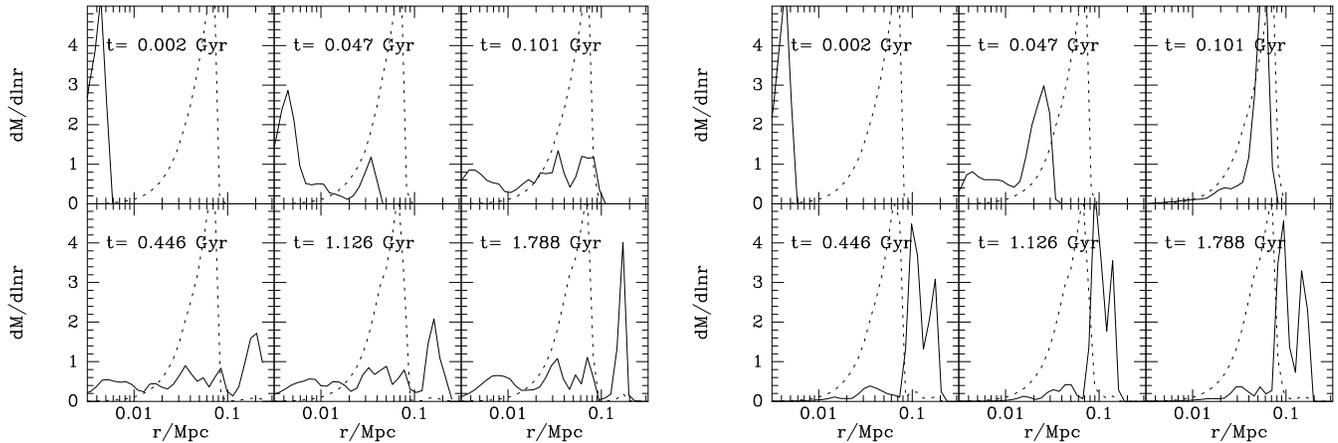

\centerline{\psfig{file=colours.2kpc.ps,width=.48\hsize}\qquad
\psfig{file=colours.3kpc.ps,width=.48\hsize}} \caption{The
distribution in log radius of material that started at $r<5\kpc$
(full curves) or at $5<r/\hbox{kpc}<77$ (dashed curves).  The
vertical scale is in units of $5\times10^{8}\msun$ for the full
curves, and $5\times10^{10}\msun$ for the dashed curves. Results
for the narrower ($r_{\rm jet}=2\kpc$) jet are shown at the left
with results for the jet with $r_{\rm jet}=3\kpc$ on the
right.\label{mixfig}}
\end{figure*}

ENZO follows the density of several dyes, each one of which
moves with the main fluid according to the usual continuity
equation. This facility enables us to explore the extent to which
an outburst moves material radially.  The full curves in
Fig.~\ref{mixfig} show the distribution at six times of a dye that
was initially distributed like the gas density interior to
$r=5\kpc$, and had zero density outside this `inner core' volume.
Data for the narrower jet are on the left. The dashed curves show
the distribution of a second `outer core' dye, which was initially
distributed like the gas density in the radial range
$5<r/\hbox{kpc}<77$ and had zero density elsewhere. The upper
middle panel for the narrower jet in Fig.~\ref{mixfig} shows that
$47\Myr$ after the jet turns on, a significant amount of
inner-core dye has moved out to the range $20<r/\hbox{kpc}<40$;
the next panel shows that when the jet turns off, the inner core
material extends out to $100\kpc$. The lower panels show that
after the jet has died, inner core material is carried out to
$r>200\kpc$ before falling back slightly in the last part of the
simulation. Similar effects are evident in the right panels for
the $3\kpc$ jet, the main difference being that this wider jet
ejects essentially all rather than most  of the inner core
material.

The dashed curves in Fig.~\ref{mixfig} show that the jet has a less dramatic
impact on the outer-core material. Nevertheless, in the left panels we see
that outer-core material is pushed out to beyond $r=100\kpc$, mostly after
the jet has switched off.  Moreover, by the end of the simulation with
$r_{\rm jet}=2\kpc$, sufficient outer-core material has flowed in to the
inner-core region for the densities of the two fluids to be essentially
equal at the centre, with the density of the outer-core material declining
much more slowly with radius than that of the inner-core material.  The
equality of the central densities is masked in Fig.~\ref{mixfig} by the
difference in the adopted mass scales.  The ability of the wider jet to
displace most of the inner core gas while affecting the outer core gas only
about as much as the narrower jet does, is consistent with the conclusion we
reached from Fig.~\ref{crossfig} that the main difference between the two
jets lies in the different quantities of plasma that they entrain near the injection
discs.

\begin{figure}
\centerline{\psfig{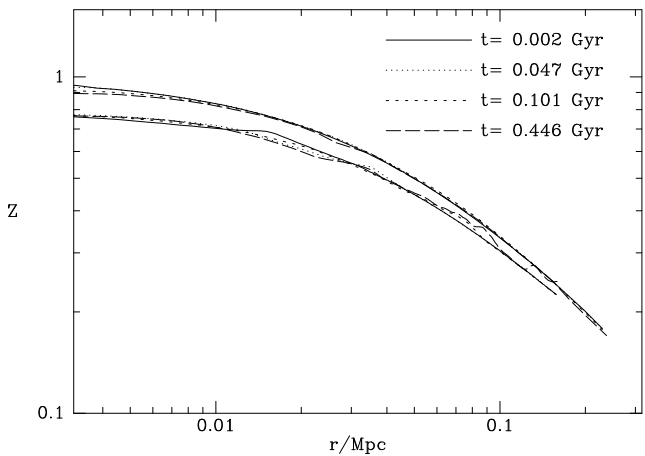}}
 \caption{Evolution of
the metallicity of the cluster gas in the simulation with the
narrower ($2\kpc$) jet. Upper curves: spherically averaged
metallicity; lower curves: metallicity derived from projected data
at points along the projected jet axis.\label{Fefig}}
\end{figure}

Fig.~\ref{Fefig} shows the evolution of the metallicity of the
IGM. This was followed by distributing a dye in the initial
configuration with a density $\rho_{\d}$ that is proportional to a
power of the main plasma density, with the power chosen such that
the `metallicity' $Z=\rho_\d/\rho$ declines by a factor 10 between
$r=0$ and $r=500\kpc$. The upper set of four curves, which shows the spherically
averaged metallicity density for the simulation with the narrower
jet, shows that the expulsion of inner-core material weakens the
central metallicity gradient slightly, but does not eliminate it.
The wider jet causes the central metallicity to decline slightly
more: to 0.857 rather than 0.888 of its original value. Naturally
changes in metallicity are largest along the line of the jets. The
lower set of curves in Fig.~\ref{Fefig} shows as a function of distance $x$ down
the jet axis the metallicity an observer
would measure along the jet if the cluster were oriented such
that the jets lie in the plane of the sky. The original
metallicity gradient is smaller than that obtained by spherical
averaging. Also at $x\sim20$ the metallicity is clearly seen to be
enhanced by uplift of material from the core.

\begin{figure*}
\psfig{file=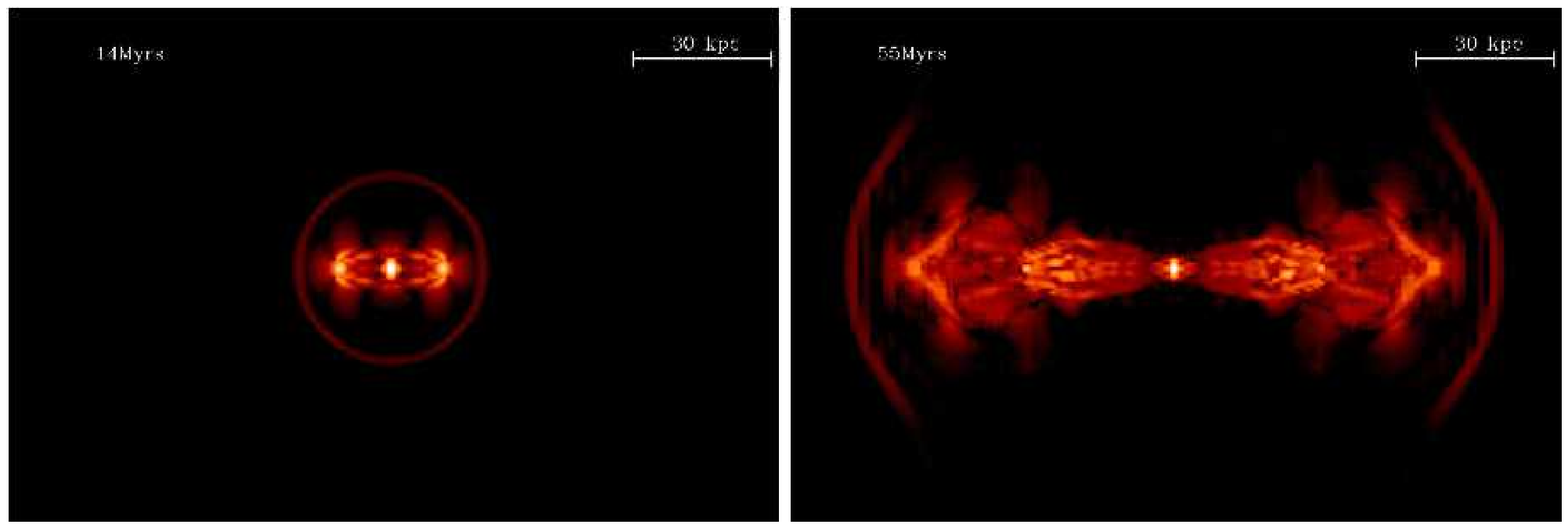,width=.96\hsize}
\psfig{file=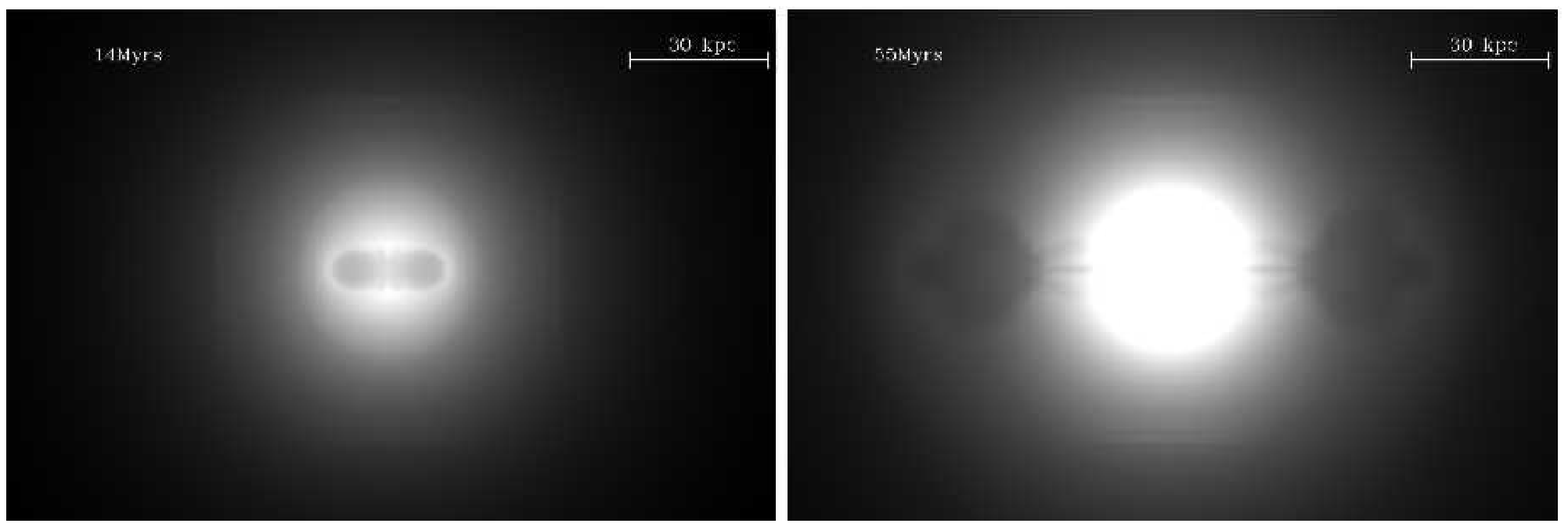,width=.96\hsize}
\psfig{file=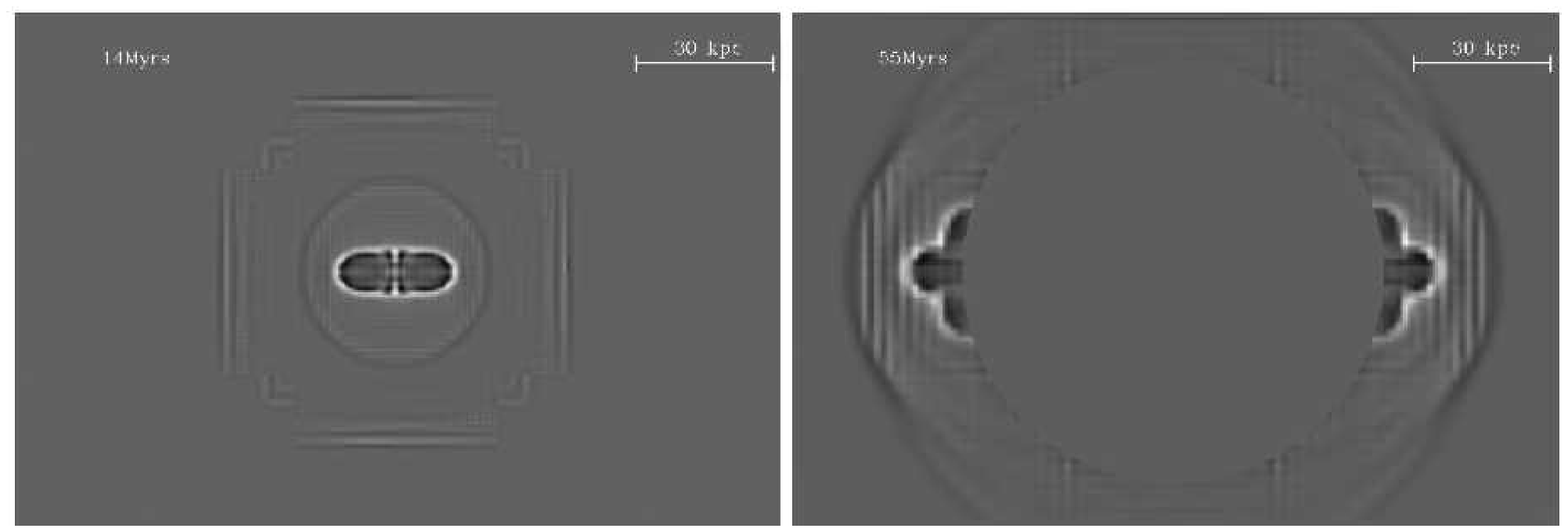,width=.96\hsize} \caption{Top row: the
divergence of the velocity field in the plane $z=0$ at $14$ and
$55\Myr$ after jet ignition shows the location of the bow shocks.
Middle row: the X-ray emissivity projected along the $z$ axis at
$14$ and $55\Myr$ show little or no trace of the shocks. Bottom
row: the shocks can be clearly seen in unsharp-masked versions of
the X-ray images of the middle row. In the $55\Myr$ case, the
central region has been blocked out to enhance the weak bow shock.
All images are for the $r_{\rm jet}=2\kpc$ case.\label{shockfig}}
\end{figure*}

The top row of Fig.~\ref{shockfig} shows the divergence of the velocity
field in the plane $z=0$ at $14$ and $55\Myr$ after ignition of the $2\kpc$
jet. At the first time shown we can see the injection disks and, nearly
connected to them, a network of bright regions that roughly mark the
boundaries of the mushrooms we encountered in the density field.  Further
out a circle of brightness delineates the bow shock as it advances into the
undisturbed IGM. At $55\Myr$ the divergence field is more complex, on
account of the emergence of internal shocks within the jets. A pair of
wedges mark the interface between ejected material and the disturbed IGM and
a little further out a bow shock appears as two sectors of a circle. The
original quasi-spherical shock is too weak to see in this plot, but the
unsharp-masked X-ray image at bottom right reveals it at about 10 o'clock
and 2 o'clock. At 3 o'clock the jet's bow shock has just penetrated the
spherical locus of the original shock.  The distance between the jet head
and the original shock has not increased between $14$ and $55\Myr$, so the
head is moving supersonically -- from Fig.~\ref{densplots2kpc} we find that
the jet head moves $37\kpc$ in the $37\Myr$ from $t=18\Myr$, so its average
speed is a factor $1.08$ times the sound speed at $3.7\times10^7\K$.

The middle row in Fig.~\ref{shockfig} shows
the X-ray
emissivity projected along the $z$ axis at the times of the top
panels. The shocks that are clearly evident in the top panels
are invisible in the X-ray data. The bottom panels show
unsharp-masked versions of the middle panels. In these images the
shocks are evident. Recently \cite{FabianEtal03} have  unsharp
masked a very long exposure of the Perseus cluster and detected numerous
weak arc-like regions of enhanced density gradient that are similar to those
seen in Fig.~\ref{shockfig}.

The right-hand panels in the middle and lower rows of Fig.~\ref{shockfig}
show X-ray cavities that are very similar to those observed in Hydra,
Perseus and other clusters
\citep{BohringerEtal93,McNamaraEtal00,FabianSET,HeinzCRB,BlantonSM03}. The
cavities in our simulations are slightly edge brightened, but rather less so
than are the observed cavities because the material at the edge is not
cooler than ambient. In part this will be because we start with an
isothermal temperature profile rather than one that drops near the centre.
It may also be the case that in a realistic cluster environment the jets
plough into dense material dredged up by a previous jet.  This would produce
a bow shock comprised of denser material, which would shine more brightly in
X-rays. It is therefore important to extend these simulations to include
multiple jet eruptions \citep{SokerBS02}.

\begin{figure}
\centerline{\psfig{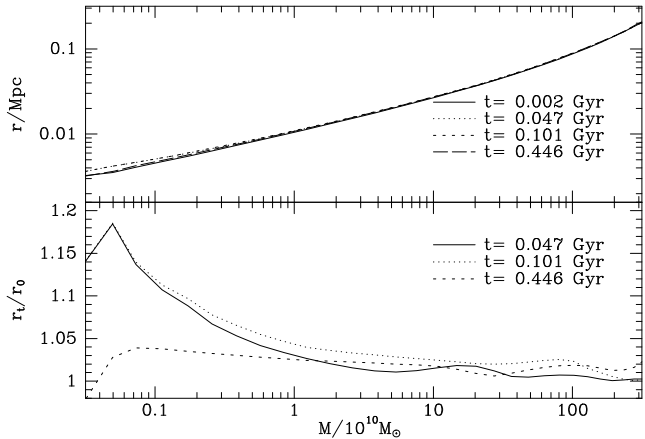}}
 \caption{$M(r)$ at
four times for the narrower jet.\label{startendA}}
\end{figure}

\begin{figure}
\centerline{\psfig{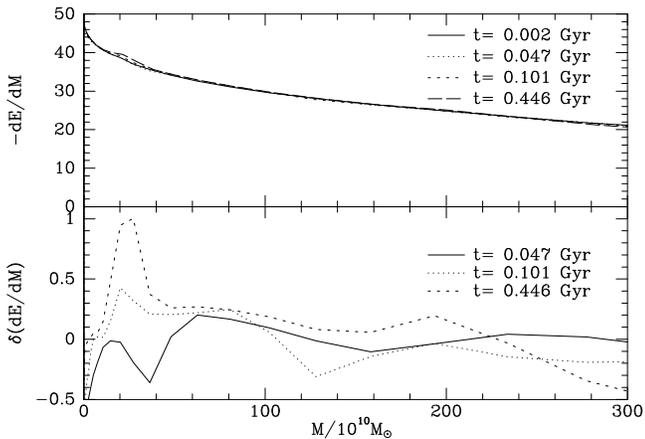}}
 \caption{$|\d
E/\d\ln M|$ at four times (units $10^{58}\,$erg per
$10^{10}\msun$) for the narrower jet. \label{startendB}}
\end{figure}

\begin{figure}
\centerline{\psfig{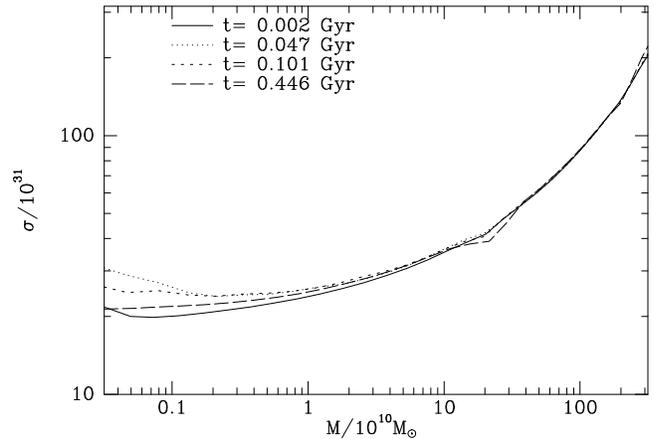}}
 \caption{Entropy
index versus mass at four times for the narrower jet.\label{sigfig}}
\end{figure}

Fig.~\ref{startendA} shows the overall impact that the jet has on
the system by comparing the radial run of mass at four times for the
narrower jet (the corresponding plots for the wider jet are  similar). The
upper panel shows the radius $r(M)$ within which mass $M$ lies.
This increases slightly at the centre. The lower panel shows this
increase in more detail by showing the ratio of the radius $r_t$
that contains $M$ at time $t$ to the radius $r_0$ that originally
contained the same mass.  Near the centre, $r_t$ is $\sim20$
percent bigger than $r_0$ so long as the jet is powered. Further
out a wave of increase in $r_t$ can be seen propagating out while
the jet is on.  By $350\Myr$ after the jet is switched off, $r_t$
is larger than $r_0$ by $\lta5$ percent at all radii.

We have seen that the jet transports cool
material from the centre and deposits it at large radii. Hence an increase
in $r(M)$ at fixed $M$ does not necessarily mean that a physical shell of
material moves outwards; the shell may in fact move inwards but its decrease
in $r$ be more than compensated by the decrease in its mass coordinate $M$.

Fig.~\ref{startendB} shows the distribution of binding energy per unit mass
as a function of mass at four times scaled such that the area under the
curves is total binding energy in units of $10^{58}\erg$. Thus the injection
of $2\times10^{59}\erg$ by the jets has the effect of shifting the mean
level of the curves vertically by only $0.1$ in the units of the vertical
axis.  The lower panel shows the difference in the binding energy per unit
mass at each enclosed mass between the given time and the moment before jet
ignition. Heating by the jet reduces binding energy and thus lowers the
curves. Around $M=40\times10^{10}\msun$ we see a pronounced dip in the curve
for $t=47\Myr$. There is a corresponding dip in the curve for $t=101\Myr$
around $M=130\times10^{10}\msun$. This increase in the radius of the dip
reflects outward propagation of energy. The decline on the extreme right in
the curve for $t=447\Myr$ implies that by this time the extra energy has
migrated to the highest mass shells included in the figure, which lie near
$r=200\kpc$, approximately the cooling radius in this cluster
\citep{DavidEtal00}.

Consider the effect of uplift of cold gas on a physical shell of matter that
lies between the radii from which the jet takes and then dumps cold gas. If
the shell is not directly disturbed by the jet, its energy will be constant,
but the value of its mass coordinate $M$ will decrease. Since in the
original configuration specific binding energy $|\d E/\d M|$ decreases with
increasing $M$, the decrease in $M$ will have the effect of decreasing $|\d
E/\d M|$ at given $M$. Hence in Fig.~\ref{startendB} uplift has the same
effect as a distributed source of heat that acts between the radii between
which the jet transports material.

The areas under the curves of the lower panel should equal the
energy injected up to each time and thus provide valuable checks
on the accuracy of the simulation. The area under the full curve
should be 9.4 and is 9.45, while the area under the other two
curves should be 20 and is 22.6 for $t=101\Myr$ and 11.4 for
$t=446\Myr$. Extremely similar values are obtained for the jet
with $r_{\rm jet}=3\kpc$. In view of the smallness of the energy
injected relative to the total energy, the agreement between the
expected and measured values at the first two times is
satisfactory. The shape of the dashed curve in
Fig.~\ref{startendB} suggests that the discrepancy at the third
time arises because the integration has not been carried out to
sufficiently large enclosed masses -- the outermost mass shell included
lies at $r\sim200\kpc$. Thus this figure indicates that the energy
injected by the jet ends near the cooling radius. A simple
calculation explains why this is so. We have seen that the jet
blows a cavity that travels at close to the sound speed and
remains an underdensity for about twice as long as the jet fires.
Hence the radius at which it becomes an overdensity and deposits
its energy is $\sim2 c_{\rm s}t_{\rm jet}\simeq180\kpc$ for our
choice $t_{\rm jet}=100\Myr$. If heating is to make good radiative
losses, the heating action of the jet needs to be concentrated at
rather smaller radii. Thus our assumed jet lifetime is likely to
be too large by a factor of a few. Since the size and energy
content of the cavities we generate are realistic, any decrease in
$t_{\rm jet}$ should be compensated by an increase in jet power.

Fig.~\ref{sigfig} shows the variation with enclosed mass of the entropy
index $\sigma\equiv P/\rho^\gamma$ in the case of the narrower jet. A prompt
increase in the entropy density at $M\lta10\times10^{10}\msun$ is evident.
Subsequently, the entropy density decreases in this zone, but remains above
its initial value. However, around $M=25\times10^{10}\msun$ the entropy
index at $t=446\Myr$ drops slightly below its initial value.  This
depression is associated with an enhancement in binding energy that is
clearly visible towards the left-hand edge of Fig.~\ref{startendB}.  From
Fig.~\ref{startendA} we see that the corresponding radial coordinate is
$r\simeq30\kpc$, which Fig.~\ref{shockfig} shows to be the extent down the
$x$ axis of a grid boundary. Consequently, there has to be a suspicion that
this feature is a numerical artifact.

\section{comparison with other work}

\cite{ChurazovEtal01}, \cite{QuilisEtal01},
\cite{BruggenKaiser01}, \cite{BruggenKCE02} and
\cite{BruggenKaiser02} injected energy only, whereas we have
injected both energy and momentum. Moreover, with the exception of
\cite{QuilisEtal01} these authors simulate only a part of the
cluster that varies in scale from the $30\times10^2\kpc$ volume of
\cite{BruggenKCE02} to a hemisphere of radius $1\Mpc$ of
\cite{BruggenKaiser01}. Only the simulations of
\cite{QuilisEtal01} and \cite{BruggenKCE02} are fully
three-dimensional, and only \cite{BruggenKaiser02} had an adaptive
grid: unfortunately, this two-dimensional simulation had slab
rather than rotational symmetry. In all these simulations energy
was injected at a fixed point that is significantly removed from
the cluster centre without any physical motivation for its
location.

The presence of the bow shock (see Fig.~\ref{shockfig}) and of the strong
turbulence in the ejected material underlines the difference between our
model and models in which `bubbles' of plasma buoyantly rise up in a
quasi-stationary manner \citep{ChurazovEtal01, QuilisEtal01,
BruggenKaiser01, BruggenKaiser02, BruggenKCE02}.  Outward motion of the
point at which the jet's energy is thermalized is clearly important for the
dynamics of cavities in our simulations, and missing from most early
discussions of bubble dynamics.

Two groups have simulated the impact of jets on cooling flows
\citep{ReynoldsHeinzBegelman01, ReynoldsHeinzBegelman02, BassonA}.  Both
groups employed a version of the ZEUS code on a single spherical grid that
excluded a sphere around the origin. In the case of Reynolds et al., the
radius of this sphere was 1 percent of that ($500\kpc$) of the outer
boundary sphere, while in the simulations of Basson \& Alexander the ratio
of these radii was $0.5$ percent and the outer bounding radius was
$2800\kpc$. Our computational domain is $630\kpc$ on a side and includes the
origin. In these simulation the PPM Riemann solver was employed, while ZEUS
uses artificial viscosity to simulate shocks.

While Reynolds et al.\ forced axisymmetry, the simulations of Basson \&
Alexander were fully three-dimensiional. In both simulations the jets were
imposed as conical inflows through inner boundary. The jets were faster and
more powerful than ours: they emerged from the inner sphere at a speed of
$\sim100\,000\kms$ and a density 100 times lower than the ambient density.
In the simulations of Reynolds et al.\ the net jet power was
$9.3\times10^{45}\erg\s^{-1}$. We estimate the power of the jets of Basson
\& Alexander to be $2.5\times10^{46}\erg\s^{-1}$. Hence, in both simulations
the jets are more powerful than ours by at least a factor 150 to 400.  The
rate of mass injection was just $1.5$ to $\sim4$ times ours. Thus the jets
were significantly less heavily loaded with momentum than ours, yet there
was enough momentum for the head, at which energy was randomized, to move
out through the cluster, rather than remaining stuck at an arbitrarily
chosen point, as in earlier work.

An important difference between the Reynolds et al.\ simulations and ours is
that in their simulations, the backflows extended all the way to the origin
for the lifetime of the jet, whereas in our simulations the backflow region
moves away from the origin at some distance from the head of the jet. This
difference is a natural consequence of the different momentum loadings of
the two jets since a higher ratio of momentum to energy increases the speed
of advance of the head relative to the speed of the backflow.

A second notable difference between the simulations is in the temporal and
spatial scales of the outbursts. In the Reynolds et al.\ simulations, the
cavities created by the jets have still not been overtaken by uplifted cold
material at 15 times the lifetime ($50\Myr$) of the jets, by which time the
cavities are $\sim1\Mpc$ from the centre. Thus in these simulations the AGN
has a major impact to the edge of the cluster. A similar situation is
evident in Fig.~1 of Basson \& Alexander. The observations suggest that the
AGN's impact should lie primarily within the cooling radius and in future
work we plan to concentrate on the part of parameter space in which this is
so. The high-power outbursts simulated by Reynolds et al.\ and Basson \&
Alexander clearly do not have promising parameters from this point of view.

All simulations except those of \cite{BruggenKaiser02},
\cite{QuilisEtal01}  and \cite{BassonA} neglect cooling, as we do.

\section{discussion}

It is now clear that AGN have a significant impact on cooling flows. The
largest question that remains unresolved is whether the best observed
systems are in an approximate steady state, or are cycling between
configurations that have significantly different central density profiles
and temperatures. Another important issue is whether the observed
cavities are driven outwards by momentum-laden jets, or simply rise
buoyantly. We have argued that the low luminosities of black holes embedded
in X-ray emitting gas, coupled with the implausibility of the ADAF model,
support the conjecture of \cite{BlandfordBegelman99} that a subrelativistic
bipolar flow is the primary output channel of these black holes. This
conclusion amounts to a {\it prima facie\/} case for momentum driving. Our
simulations explore the regime of strong momentum driving, and provide a
counterpoint to simulations in which energy alone is released at some
point in the flow.

We have modelled the impact on a
cooling flow similar to that in the Hydra cluster, of an outburst
by an AGN that lasts $100\Myr$ and has mechanical power
$6\times10^{43}\erg\s^{-1}$ delivered by opposing jets. The jet
power is smaller than the X-ray luminosity of the Hydra cluster,
so to achieve a quasi-steady state a series of more powerful outbursts
would have to follow one another. We have sought to gain physical
insight into the overall response of the cooling flow to energy
input by jets by suppressing both subsequent outbursts and
radiative cooling.

The jets are created within the simulation by an algorithm that avoids the
imposition of any boundary conditions near the cluster centre. The speed and
early development of the jets depends on the width of the base of the jets
because wider jets entrain more ambient material at their base. The gross
features of jet action are independent of the assumed jet width, however.
Cavities of very hot plasma grow around the heads of the jets, and these
move supersonically outwards.  Ahead of the cavities run two cap-like bow
shocks, one over each jet head, which eventually overtake the weakening
spherical shock front that is generated as the jets ignite. The cavities,
which advance supersonically, are evident in simulated raw X-ray images, but
the bow shocks can only be detected in the X-ray data when they are
unsharp-masked.

A turbulent vortex that contains a significant quantity of entrained and
uplifted material trails each cavity. After the jet has turned off the
cavity weakens and slows relative to the denser vortex. About $100\Myr$
after the jet turned off, the vortex fills and overtakes the cavity, which
is thus replaced by a region of outward-moving, cool, overdense gas.
Eventually this over-density falls back inwards. Many studies (e.g.\
\cite{Alexander02} and references therein) have pointed out that an AGN
outburst will lift significant quantities of low-entropy material out of the
cluster core. However, they have discussed this process in the context of
shock-acceleration of material ahead of a rising cavity. It is not clear
that shock-acceleration causes the dense following vortex that we see; this
structure is probably a part of the global circulation pattern that the jets
excite, in which gas flows in perpendicular to the jet axis and out
parallel to the axis.

We have used dyes to assess the importance of turbulent mixing.
Much or most of the material that was originally in an inner core
of radius $5\kpc$ is ejected to radii as large as $200\kpc$.
Material that originally lay between $5$ and $77\kpc$ moves in
part inward to replace the ejected inner-core material, and in
part outwards. Most stays put, with the result that an original
metallicity gradient in the IGM is only slightly weakened.

There is a small overall inflation of the IGM: while the jets are
still firing, the radius at a given value of the mass coordinate
$M$ increases by $\sim10$ percent within a few kiloparsecs of the
centre. The long-term increase in radius of material within
$\sim100\kpc$ is only of order a percent. The entropy index of the
innermost $\sim10^{11}\msun$ increases by up to a few percent.
Most of the energy injected by the jets ends up around the cooling
radius or beyond because the distance that the cavities move is of
order $2c_{\rm s}t_{\rm jet}$. Heating could more readily offset
cooling if a smaller value of $t_{\rm jet}$ and a larger value of
the jet power were used.

We now return to  the questions posed in Section 3:
\begin{enumerate}

\item The outflow from an accreting black hole must carry momentum as well
as energy. Most previous studies have neglected the momentum flux because
they have been directed at the effects of ultrarelativistic jets, that
probably consist of a very light electron-positron plasma. We have stressed
the likelihood that any such jet lies at the core of a massive,
subrelativistic bipolar flow that comes from the accretion torus rather than
the black hole's ergosphere. Our assumption of a non-negligible momentum
flux enables our simulations to be self-consistent in a way that the
energy-only simulations are not: in our simulations the point at which the
jet's ordered kinetic energy is randomized is self-consistently determined,
whereas in the simulations of \cite{BruggenKaiser02} and others the point
at which energy is injected is arbitrarily chosen. Moreover, cavities that
are pushed out into the ambient medium are more stable than ones that merely
rise buoyantly and we tentatively infer from the morphology of the Virgo A
radio source that the data favour stable cavities.  Specifically,
Fig.~\ref{M87fig} shows that a feature in the radio-continuum map of Virgo A
in \cite{OwenEilekKassim00} closely resembles a feature in one of our
simulations. This resemblance cannot count as hard evidence for momentum
driving because the quantities plotted in the two cases are very different.
But our impression is that no feature in the energy-only simulations so
closely resembles the observed feature because the simulated features are
too prone to fragmentation.

\begin{figure}
\centerline{\psfig{file=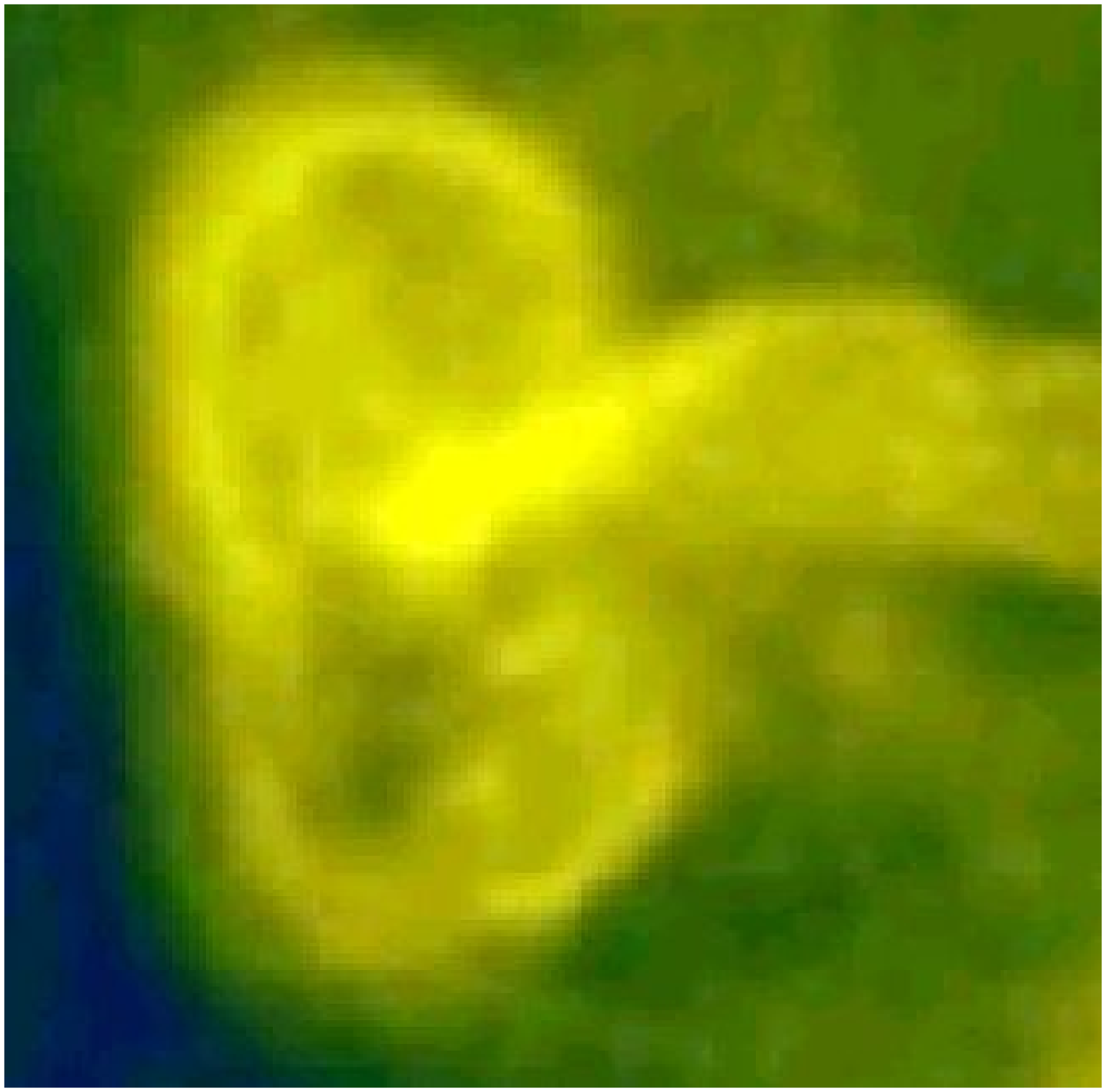,width=.45\hsize}\
\psfig{file=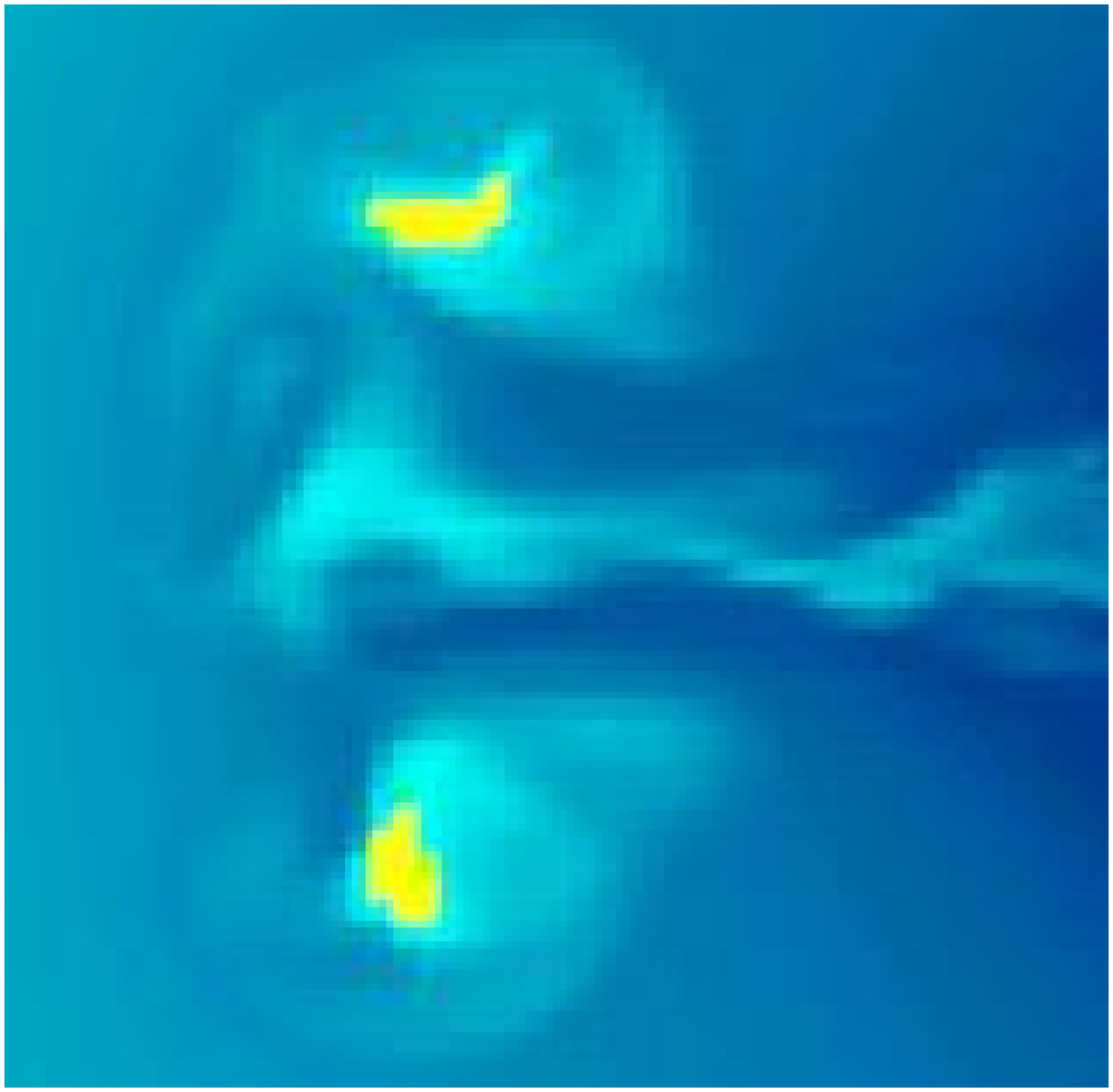,width=.45\hsize}}
\caption{Left: a piece of the radio-continuum map of Virgo A by
\citep{OwenEilekKassim00}. Right: a piece of the map of specific entropy
in the  plane $z=0$ from one of our simulations.
\label{M87fig}}
\end{figure}

\item The jet's energy offsets the effects of radiative cooling  in several
ways: 1) gas at the very centre is pushed up high, where it mixes in with
higher-entropy material and is heated by dissipation of the system's
strongly excited g-modes; 2) gas slightly further out is first shock
heated and then adiabatically compressed as it sinks inward to take up the
volume vacated by ejected gas; 3) gas at all radii is heated by being
turbulently mixed with the extremely hot shocked jet material; 4) the
radio map of \cite{OwenEilekKassim00} strongly suggests that in Virgo
shocked jet material is distributed throughout the inner $\sim40\kpc$. If
the jet consisted primarily of ultrarelativistic material, it would not
follow that the thermal plasma had gained significant heat from the jet
because the coupling between ultrarelativistic particles and thermal plasma
is weak. If we accept the likelihood that the observed ultrarelativistic jet
is sheathed in a subrelativistic wind, it does follow from the radio map
that turbulent mixing will have heated the thermal plasma within
$\sim40\kpc$ of the cluster centre.

\item The rate of jet heating can be estimated as the energy
within the observed cavities, divided by cavity lifetime
$\sim2t_{\rm jet}$. Our conclusion that cavities survive for about
twice the time it takes to inflate them is consistent with the
observation of two pairs of cavities in clusters such as Perseus,
one pair being radio bright and the other a pair of ghost
cavities. Since cavities move at the speed of sound, we can
determine their lifetimes from how far out we find them. For
Perseus we thus infer $t_{\rm jet}=15\Myr$, and a heating rate
$P=1.4\times10^{44}\erg\s^{-1}$ per $7\kpc$-radius cavity. This is
close to but less than the X-ray luminosity from within the
cooling radius. So are cooling flows in steady states? We think
not. First,  it is unlikely that the black hole's output can be
closely matched to the radiative cooling rate because the
characteristic timescales of the black hole and the cooling flow
are so discrepant. Second, as \cite{Motl} and \cite{KayST03} have
stressed, clusters experience substantial infall of gas, and a
significant quantity almost certainly reaches the centre without
being shock heated to the virial temperature. It is likely that
the gas observed in molecular form
\citep{Donahue_etal00,Edge01,EdgeEtal02} and as H$\alpha$-emitting
filaments \citep{Lynds70,CowieEtal83,Conselice01} is just such gas
rather than gas that has cooled in the flow: as
\cite{SparksEtal89} have stressed, dust in the filaments would be
unlikely to have perfectly normal Galactic extinction properties
if the gas had condensed from the hot phase.  Moreover, the
filaments would not have their highly disturbed morphology if they
had condensed from the flow.  X-ray data from {\it Chandra\/} show
that the cold gas is in thermal contact with the X-ray emitting
plasma \citep{FabianCrawford03}, and it is probably being
evaporated by the latter. If so, this activity will be a
significant drain on the energy resources of the cluster gas,
making it likely that, notwithstanding current AGN heating,
clusters such as Perseus are drifting towards a cooling
catastrophe.

\item After an outburst, hydrostatic equilibrium is approached
after several Brunt--Vaisala times. From Fig.\ref{BVfig} we see
that this timescale is very long, being $\gta1\Gyr$ at $r=200\kpc$
and still larger further out. Consequently cooling-flow clusters
should show significant deviations from hydrostatic equilibrium at
most radii, and several different AGN outbursts will have an
impact at the current time.

\item Raw X-ray images are expected to show cavities but not
shocks, even though the cavities move supersonically. If X-ray data of
exceptionally high signal-to-noise are available, unsharp-masking will
reveal the shocks. Cavities live for about twice as long as the jet that
inflates them, so there should be 2--4 cavities visible at any given time.
We do not simulate the production of synchrotron-emitting electrons, but in
view of the short lifetimes of these particles ($\sim40\Myr$ for Lorentz
factors $\gamma\sim10^4$), one would not expect cavities to be bright
synchrotron sources for long after their parent jet died. Hence our
simulations suggest that a significant fraction of cavities should be
radio-invisible `ghost' cavities.

\item Although Fig.~\ref{sigfig} shows that the specific entropy
of matter at the centre is increased by the AGN,
Fig.~\ref{startendB} suggests that most of the energy injected by
an outburst ends up near or beyond  the cooling
radius. This result follows from the facts 1) that much of the
jet's energy goes into inflating a cavity which moves mildly
supersonically, and 2) that we have assumed that the jet fires
for $100\Myr$ and find that the cavity survives for a further
$100\Myr$ after the jet expires. Travelling at $\sim1000\kms$ for
$200\Myr$, the cavity moves out to $\sim200\kpc$ before depositing
its energy.

If one accepts that averaged over gigayear times, heating by AGN offsets
radiative cooling, then the heating ought to be concentrated well within the
cooling radius, which is $\sim200\kpc$ in the Hydra cluster
\citep{DavidEtal00}.  Moreover, \cite{OwenEilekKassim00} find that in Virgo
the synchrotron emission has a sharp outer boundary at $\sim60$ percent of
the cooling radius. This observation suggests that in Virgo no cavity has
reached the cooling radius, and \cite{SokerBS02} remark that observed radio
jets do not propoagate to large distances in cooling-flow clusters. We
tentatively conclude that in our simulations the cavities move to
unrealistically large radii.  This deficiency would be resolved by halving
the lifetime of the outburst. Since the energy in our cavities is about
right, this halving of the lifetime would have to be offset by a doubling of
the jet power.

\item A crude picture of the impact of jets on the ICM is that the
jets entrain a substantial body of gas from the very centre, heat it and
transport it to rather large radii, where it mixes in with ambient material.
Material that originally overlay the ejected gas then sinks inwards to take
its place.  In this picture the curves of metallicity versus radius in the
left panel of Fig.~\ref{Fefig} would simply stretch to the left at constant
metallicity because the gas from the centre would have a negligible effect
on the metallicity of the much larger quantity of gas with which it mixed at
large radii. In this case the decline in the central metallicity gradient
would be very small.  Fig.~\ref{Fefig} indeed shows that, although an
outburst flattens the metallicity gradient, the effect is confined to
$r<20\kpc$ and is small. Ongoing
stellar evolution would have a countervailing tendency to increase the
metallicity gradient. Thus there does not seem to be a conflict between
heating by jets and the existence of metallicity gradients near cluster
centres.

\item Since Fig.~\ref{BVfig} shows that the dynamical time of the
ICM exceeds $200\Myr$ at all radii, the large-scale structure of the ICM
will be unaffected by fluctuations in the AGN's power on short timescales.
However fluctuations on shorter timescales may be important.  Of particular
interest is the timescale on which the jet channel closes up after the jet
turns off, since a jet that is quiescent for this time or longer can
probably not reactivate the cavities that it was formerly inflating. Hence
new cavities will be formed when the jet comes back online. Our simulations
suggest that the channel fills up more quickly than they are cut, but
shorter timescales are conceivable, especially if the ICM is rotating and
the jet does not fire along the axis of rotation.

\end{enumerate}

Although great strides have been made in our understanding of cooling flows
in the last few years, many key questions remain open. The cooling flow
phenomenon is an extremely important one both because of its connection with
galaxy formation, and because of the insights it gives to the dynamics of
AGN. Simulations will have an important role to play in understanding
cooling flows because they are inherently dynamical systems.  Although a
certain amount of progress can be made with axisymmetric simulations, the
chaotic and turbulent nature of cooling flows makes essential
three-dimensional simulations such as those we have presented here.
Limitations of the simulations that arise from only limited spatial
resolution are a worry, particularly in the three-dimensional case. For this
reason we have chosen to focus in this paper on just two simulations, which
are highly artificial in that they exclude radiative cooling. These
simulations have clarified a number of numerical issues and provided a
foundation for further work. We are currently experimenting with
shorter-lived, more powerful jets and with simulations that include cooling
and rotation of the cluster gas.
Animations of the simulations described here and a number of other
simulations can be viewed at www.clusterheating.org.

\section*{Acknowledgments}

Work by HNO was made possible by funding from the Norwegian
Research Council. JJB thanks Merton College for support. AS
acknowledges the support of a fellowship from the UK Astrophysical
Fluids Facility. We thank Katherine Blundell for valuable comments
on an early draft of this paper.

\end{document}